\documentclass{PoS}

\usepackage{url}

\title{Proving the outstanding capabilities of Imaging Atmospheric Cherenkov Telescopes in high time resolution optical astronomy}

\ShortTitle{Proving the outstanding capabilities of IACTs in high time resolution optical astronomy}

\author{T. Hassan$^{1}$, M.~K.~Daniel$^{2}$, 
for the VERITAS Collaboration\footnote{For collaboration list see PoS(ICRC2019)1177} \footnote{http://veritas.sao.arizona.edu}\\
       $^{1}$ Deutsches Elektronen-Synchrotron (DESY), Platanenallee 6, D-15738 Zeuthen, Germany\\
       $^{2}$Center for Astrophysics $|$ Harvard \& Smithsonian, Fred Lawrence Whipple Observatory, Amado, AZ 85645, USA\\
      E-mail: \email{tarek.hassan@desy.de}, \email{michael.daniel@cfa.harvard.edu}, 
       }
       
\abstract{Imaging Atmospheric Cherenkov Telescopes (IACTs) are very-large telescopes designed to detect the nanosecond-timescale flashes produced within extended air showers. Because IACTs are sensitive to the Cherenkov light (UV/blue) and use photodetectors with extremely fast time responses, they are also able to perform simultaneous optical observations. The large reflecting areas of these telescopes (larger than 100 m$^2$) makes them well-suited to studying fast optical transient phenomena with timescales ranging from seconds to milliseconds to nanoseconds, and the unique optical design provides a wide field of view monitoring capability with a modest point spread function. VERITAS, with its recently upgraded PMT current monitoring instrumentation, was able to provide the first detection of asteroid occultations with an IACT, resulting in the highest angular resolution measurements for stellar diameters ever taken in the visible band range. Here we explore the feasibility of using this technique to significantly expand the number of stars with directly measured stellar radii, usable for population studies to test stellar evolution modelling or transiting exoplanet radius measurements. A single observatory with a high-speed visible-band photometer with a sensitivity reaching the 13$^{th}$ magnitude could increase the number of directly measured K stars diameters by 50\%.}

\FullConference{36th International Cosmic Ray Conference -ICRC2019-\\
		July 24th - August 1st, 2019\\
		Madison, WI, U.S.A.}

\begin{document}

\section{Introduction}

Over the last decade, the current generation of IACT arrays has proven to be the most sensitive instruments to study the very-high-energy gamma-ray universe (from about 100 GeV up to 50 TeV)\cite{Weekes96}. These arrays of multiple large-diameter ($>$ 10 m) telescopes are designed to simultaneously image the nanosecond-timescale Cherenkov flashes produced within extended air showers, in order to reconstruct the impinging primary gamma-ray direction and energy. In order to detect these flashes, mainly emitte in the ultraviolet and blue, IACTs use photodetectors with extremely fast time responses, sampled at up to GHz frequencies \cite{VERITAS_camera}.

\begin{figure}
\centering\includegraphics[width=0.38\linewidth]{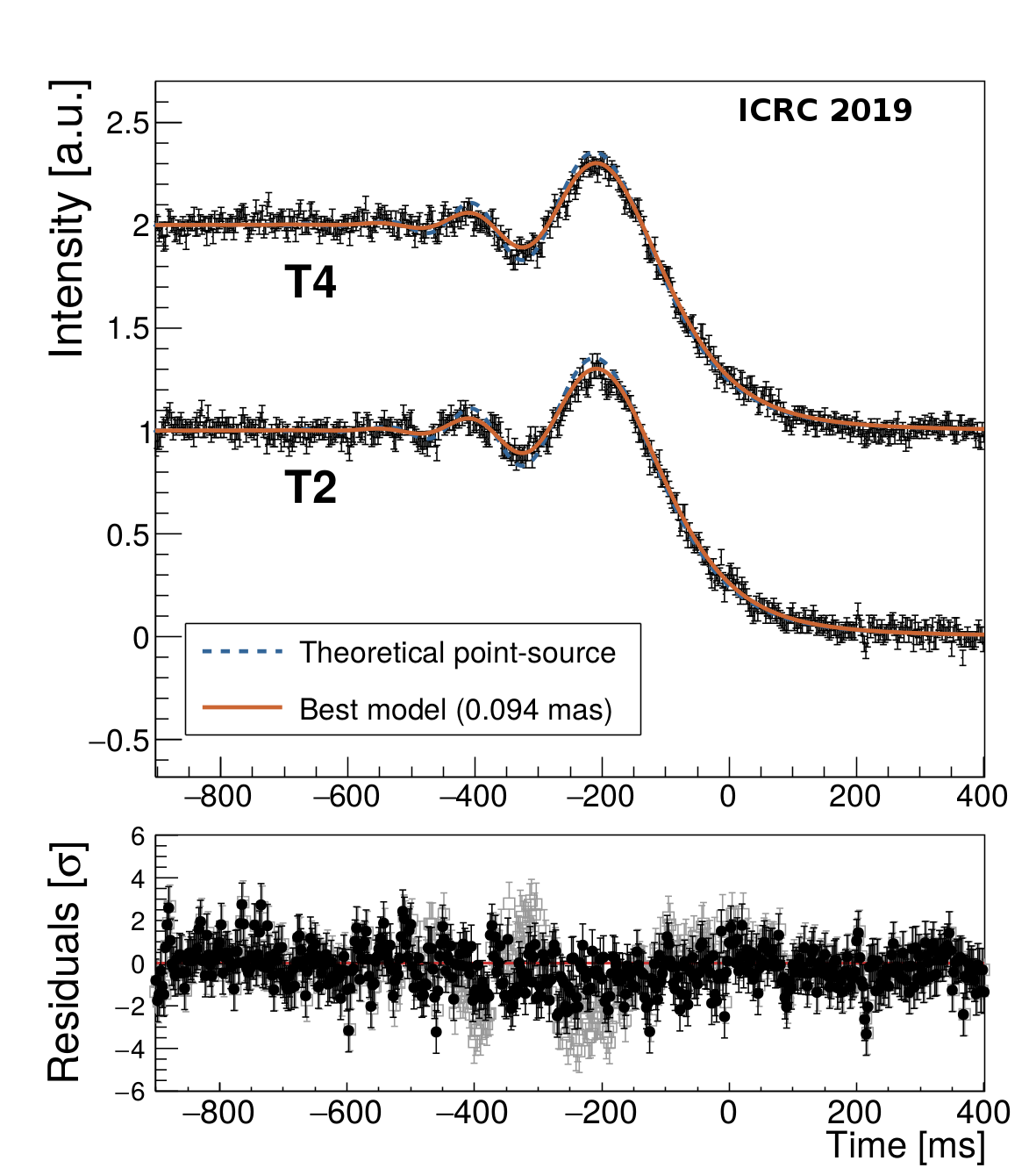}
\centering\includegraphics[width=0.38\linewidth]{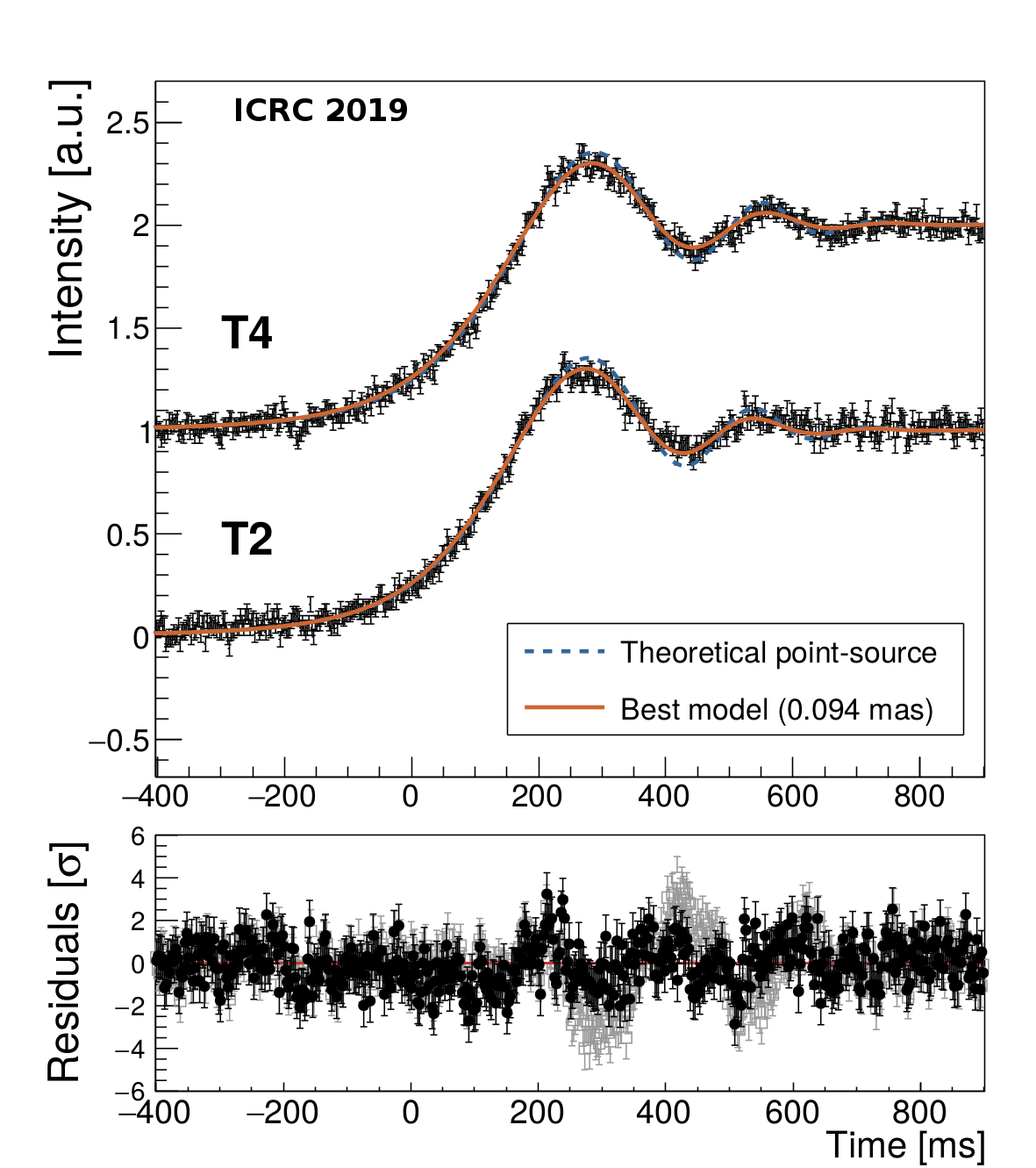}
\caption{\label{fig:Penelope_lightcurves}
The light curves of the ingress (left) and egress (right) of the (201) Penelope / TYC\,278-748-1 occultation \cite{natureAstronomy}, with the best-fit diffraction pattern (red line) and theoretical point-source model (dashed blue line). 
The combined (averaged) residual with respect to the point-source (grey empty squares) and best-fit (black filled circles) models are shown in the bottom panels.}
\end{figure}

In addition to their capabilities for indirect gamma-ray detection, these telescopes are also able to perform direct measurements as high-precision photometers \cite{lacky} in visible wavebands. These capabilities have generally not been exploited due to the modest optical quality of their mirror surface. But recent efforts have shown that these telescopes, with minor additions to their hardware \cite{centralPixel}, are able to operate as competitive very-fast optical detectors in the millisecond regime \cite{magicSensitivity}. The recent observations of 2 asteroid occultations by the Very Energetic Radiation Imaging Telescope Array System (VERITAS) observatory led to the highest angular resolution measurement for stellar diameters ever taken. This was possible by detecting and analysing the diffraction pattern surrounding the asteroid's shadow \cite{natureAstronomy}. These results prove the outstanding capabilities of these telescopes in high time resolution optical astronomy. Figure \ref{fig:Penelope_lightcurves} shows, as an example, the lightcurves of the ingress and egress of (201) Penelope asteroid occulting TYC\,278-748-1.

\begin{figure}
\centering\includegraphics[width=\linewidth]{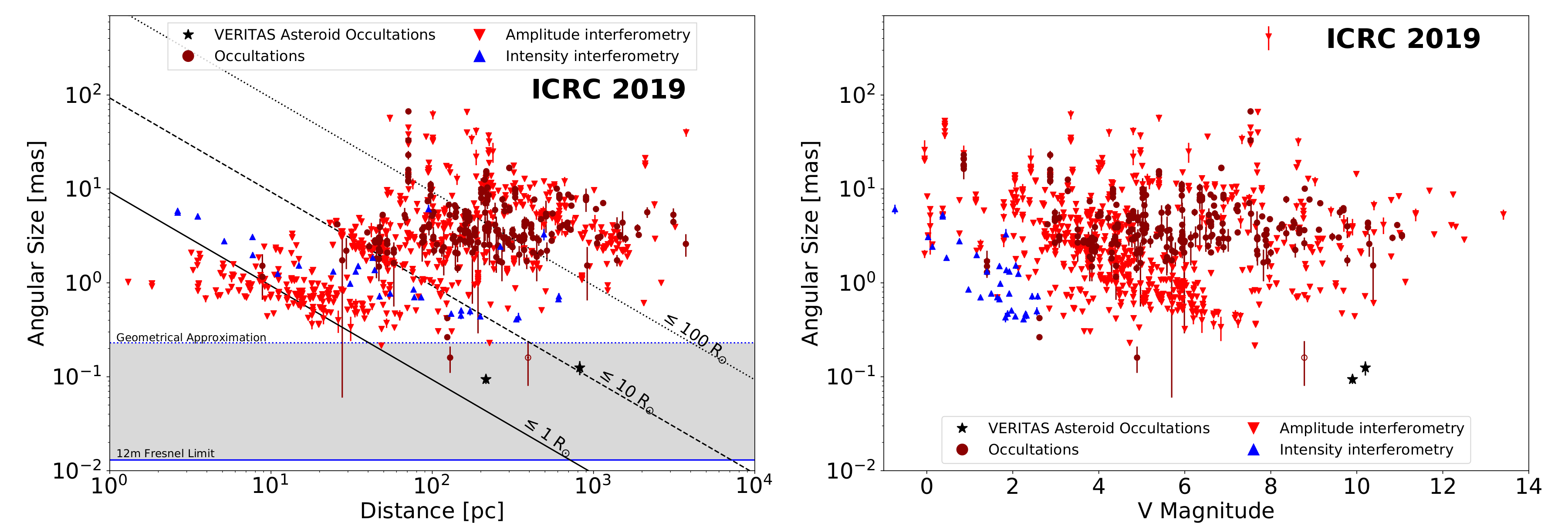}
\caption{\label{fig:starsize_directly_measured}
The angular size as a function of distance for all stars with direct angular size measurements \cite{JMDC}. 
The VERITAS asteroid occultation measurements are marked with black stars; all other occultation measurements by dark red circles; amplitude interferometry measurements by downward pointing red triangles; intensity interferometry by upward blue triangles. Note marker colors were chosen to also indicate the optical wavelength at which these measurements are performed.
The solid, dashed and dotted black lines respectively show the expected angular size for a 1, 10 and 100 solar radius star as a function of distance. 
The blue solid line gives the theoretical limit for discriminating between a point-like source and a resolved star by its Fresnel diffraction and the blue dashed line the region where the diffraction pattern disappears for a geometrically resolved star. 
\textit{Right}: As before, but for angular size as a function of apparent magnitude. Errors for this work are the 68\% confidence level, all others are taken from the respective catalogue entry \cite{JMDC}.}
\end{figure}

The direct measurement of stellar angular diameters (shown in Fig. \ref{fig:starsize_directly_measured}) have been made mainly through 3 different techniques: 
\begin{itemize}

\item Lunar occultations: when using the Moon as the occulting object, the diffraction pattern fitting technique has been used to directly measure the size of stars down to the $\sim$1 mas scale \cite{richichi}. This technique is usually limited toward the red end of the optical spectrum ($\lambda>600$\,nm) where background light from the Moon is minimised.

\item Amplitude interferometry \cite{CHARA} observations allow the direct measurement of stars (with imaging capabilities), but are also largely limited to the redder end of the spectrum due to the atmospheric scintillation noise. Scintillation limits the ability of the interferometer to correct the optical path length to the necessary fraction of a wavelength for telescope separations more than $\sim100$\,m (which equates to angular scales of $\geq$ 200 \,micro-arcseconds).

\item Intensity interferometry \cite{SII} is an alternative method free from scintillation noise, able to extend into the blue end of the optical spectrum. This technique requires very large mirror surfaces and is intrinsically limited to only the measurement of bright, hot sources (historically m$\leq3$, T$\geq 10,000$\,K, but more sensitive instruments are in development \cite{VTSSII, CTAWP}). 

\end{itemize}

In this contribution we will review the capabilities of VERITAS as a high-speed visible-band photometer and discuss the feasibility of using asteroid occultations as a novel technique to perform sub-milliarcsecond resolution direct measurements of stellar angular diameters with angular resolution and stellar magnitude sensitivity currently not accessible to interferometers. Section \ref{sec:asteroidOccultations} will focus on describing the asteroid occultation diffraction fitting technique and will evaluate the number of detectable events expected from a fixed site. Section \ref{sec:opticalVeritas} will describe VERITAS capabilities as a system of high-speed visible-band photometers and the hardware setup used. In section \ref{sec:improvements} we propose several hardware upgrades that could be implemented to improve the accuracy and sensitivity of these measurements. Finally, we devote section \ref{sec:conclusions} to discuss our results.

\section{The asteroid occultation diffraction fitting technique}
\label{sec:asteroidOccultations}

When asteroids pass in front of a star as viewed on the celestial sphere, their shadows are projected across the Earth's surface. If the occulted star is observed with a fast optical detector located within the path of the cast shadow, a rapid drop/rise in the light intensity will be measured during ingress and egress. Such measurements have been classically used to study in detail the shape and nature of distant asteroids \cite{haumea}. Recent measurements with VERITAS \cite{natureAstronomy} proved that these occultations are also a viable method to effectively measure star diameters with unprecedented resolution.

\begin{figure}
\centering\includegraphics[width=0.4\linewidth]{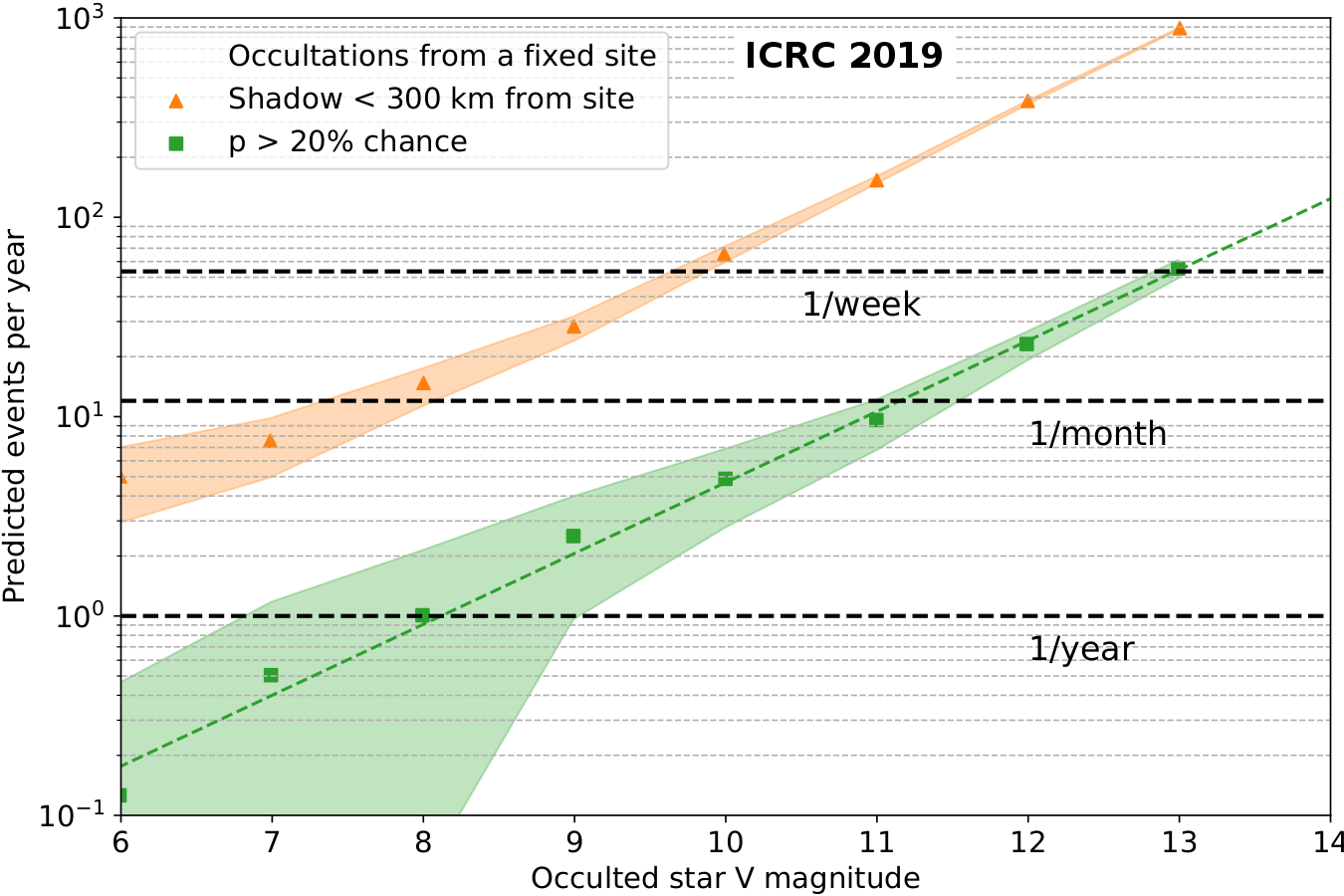}
\centering\includegraphics[width=0.48\linewidth]{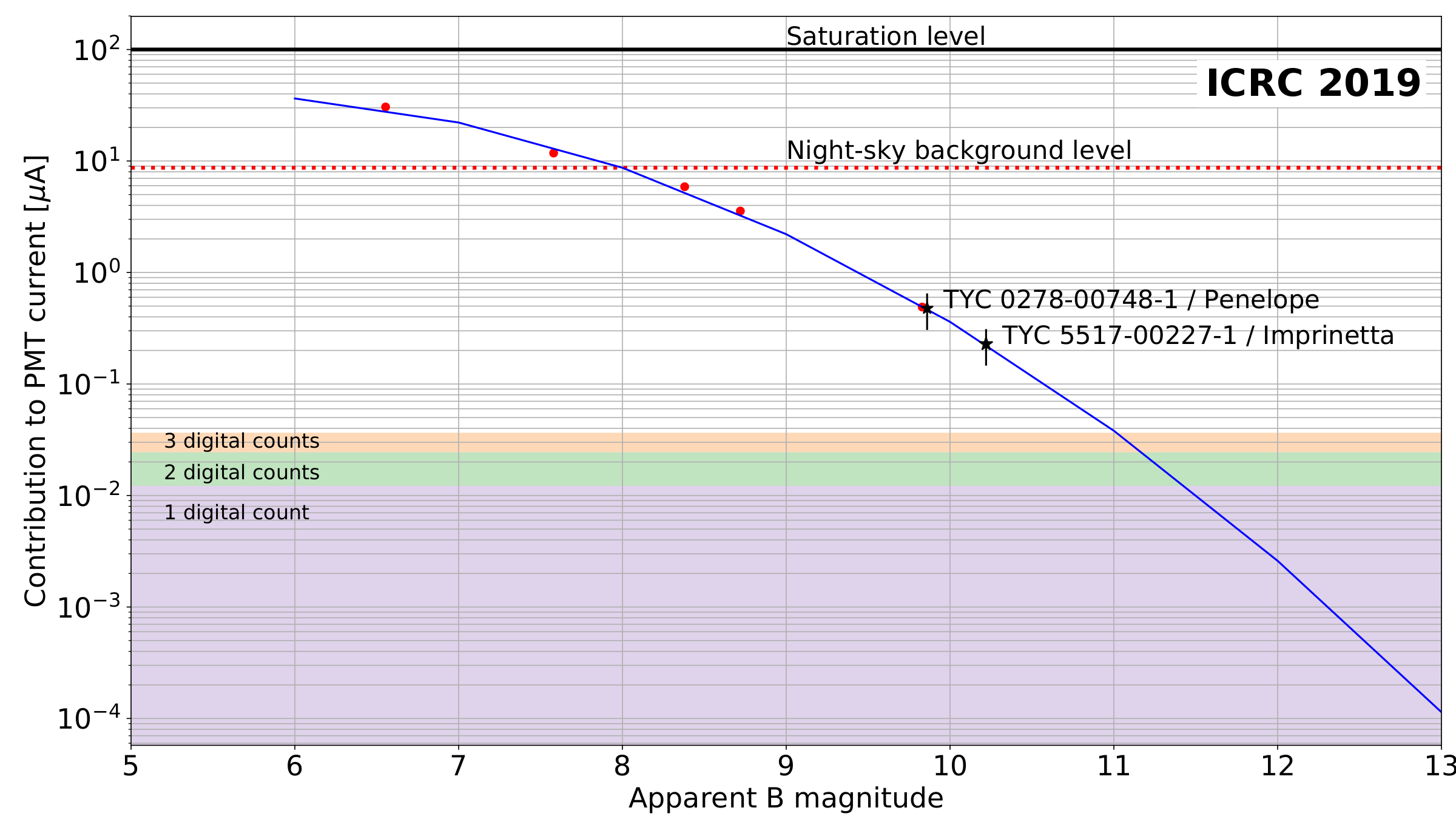}
\caption{\label{fig:predicted_occultations}
\textit{Left}: Number of predicted asteroid occultation events per year as a function of the occulted star V magnitude observable from a fixed location. Orange dots refer to any asteroid shadow passing less than 300 km away from the site while green squares refer to the number of predicted occultation events with a probability of the shadow actually passing over the site (detectable occultation) larger than 20\%. It was calculated by using the Occult software \cite{occult}, using the Gaia DR2 catalogue \cite{gaiadr2} of star positions (m$_V$ < 14), averaging the predicted occultation rates between 2012 to 2019 at the Fred Lawrence Whipple Observatory. \textit{Right}: VERITAS Enhanced Current Monitor relation between pixel current and equivalent star apparent magnitude. The current of the night-sky background and saturation levels are shown. 
}
\end{figure}

The asteroid occultation diffraction fitting technique, as compared with other methods to directly measure stellar diameters, is not limited to the brightest stars in the sky. This allows the measurement of stars with unprecedented low brightness.
In addition, by measuring the diffraction pattern produced by main-belt asteroids (at a distance an order of magnitude farther away than the Moon) the attained resolution may reach the tens of microarcsecond scale. On the other hand, the method has a limitation in that target stars cannot be chosen, and observations rely on asteroid occultation predictions which typically have high uncertainties. The number of stellar occultations predicted to be observable from a fixed observatory site as a function of the occulted star magnitude is shown in Figure \ref{fig:predicted_occultations} (left). Specifically, IACTs will be able to measure stellar diameters down to magnitude $\sim$11/13, while other high-speed cameras operating in the visible waveband (e. g. HiPERCAM \cite{hipercam}, Visitor Instrument on the Gran Telescopio Canarias) would be able to properly measure occultations fainter than 14th magnitude. 

As discussed in \cite{occultation_predictions}, the GAIA mission will significantly improve the precision of asteroid occultation events predictions. Consequently, the predicted rates shown here should be considered rather conservative. The accuracy that is currently possible with large main-belt asteroids of 100 km diameter will soon be achieved for much smaller (and numerous) 15 km asteroids.

\section{VERITAS as a high-speed optical telescope}
\label{sec:opticalVeritas}

VERITAS is a system of four 12\,m diameter segmented reflectors equipped with a camera containing 499 photomultiplier tube (PMT) pixels. Each pixel subtends a 0.15$^\circ$ field of view closely matching the optical point spread function of the reflector \cite{VERITAS}. 
In order to significantly improve VERITAS capabilities as a millisecond high-speed optical photometer, each VERITAS camera was instrumented with a commercial DATAQ DI-710-ELS DC voltage datalogger to monitor the DC current of up to 16 pixels per camera. The DATAQ logger records the DC light level in each pixel with 14-bit resolution and sampling rates up to 4,800\,Hz.
This datalogger and the Cherenkov data acquisition can be used for simultaneous optical and gamma-ray coverage. 

The main source of noise for fast optical measurements 
is the scintillation noise added by atmospheric turbulence \cite{Scintillation}. For a telescope of diameter $D$ and a sampling time $t$, the intensity fluctuations from scintillation noise scale as $\Delta{I}/I \propto D^{-2/3}/\sqrt{t}$. This means the 12\,m VERITAS telescopes, instrumented with millisecond sampling, should have at least 20 times lower noise level than a standard asteroid occultation telescope. These are typically 50\,cm telescopes equipped with a high frame rate ($\sim$60\,Hz) video camera. 

In the case of IACTs, their high-speed sensitivity in visible wavebands is mainly limited by the (irreducible) shot noise in the PMT DC current. When considering the large field of view IACTs integrate over a single pixel, 0.15$^\circ$ (0.1$^\circ$) in the case of VERITAS (MAGIC), the integrated night sky background contribution is equivalent to the flux expected from V $\sim$ 8 (10) stars. The choice of data acquisition system may also limit the minimum signal an IACT is able to identify with high confidence over a brief period of time. As shown in figure \ref{fig:predicted_occultations} (right) the current limiting magnitude of the VERITAS setup, set by the resolution of the datalogger, is about $\sim$ 11/12 mag in the B band over millisecond timescales.

\section{Possible hardware setup improvements}
\label{sec:improvements}

\begin{figure}
\centering\includegraphics[width=0.9\linewidth]{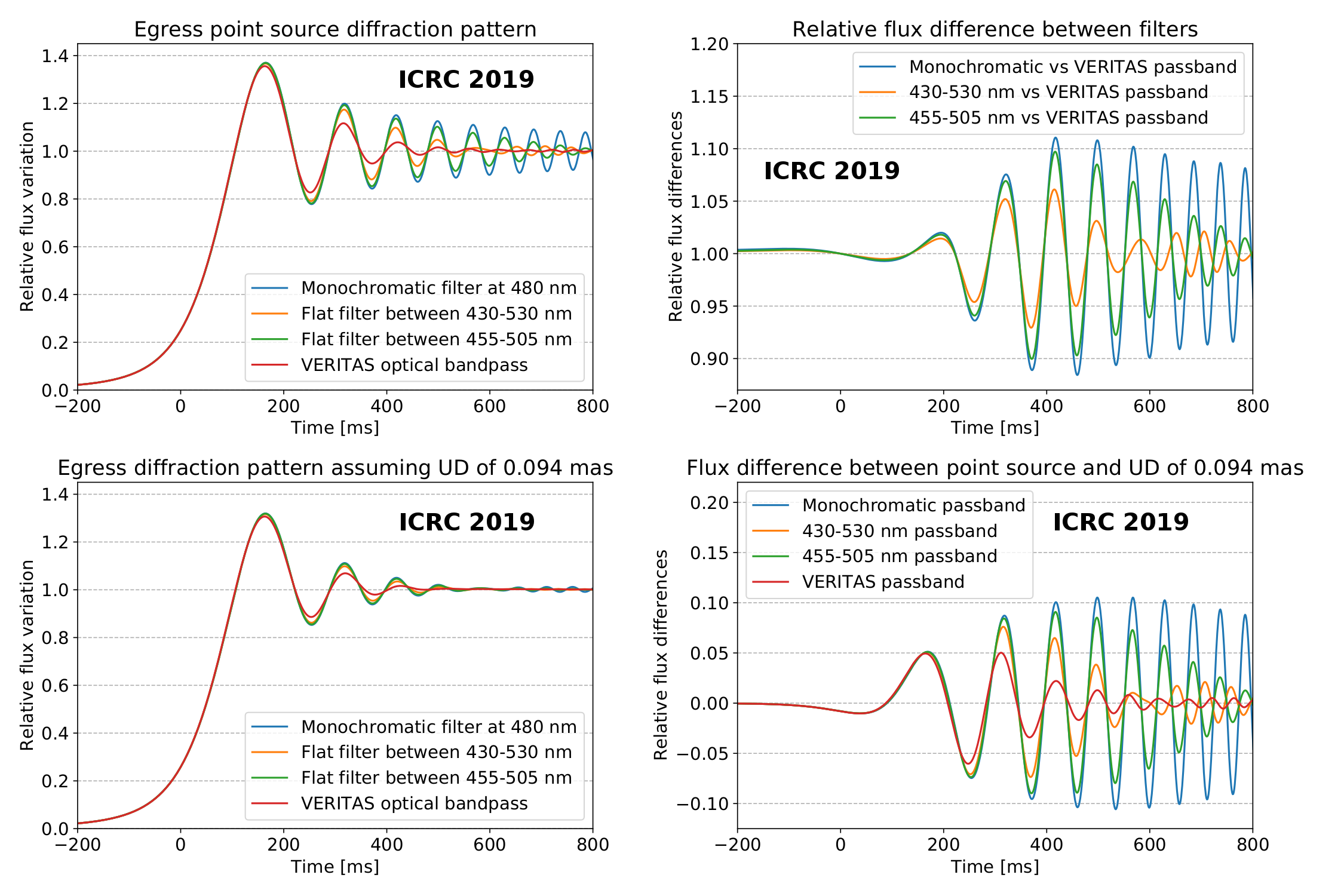}
\caption{\label{fig:narrowing_bandpass}
Differences between the diffraction pattern produced over different optical bandpass when assuming a point-like source (top figures) and an extended uniform disk source (bottom figures). Note the bottom-right figure shows the relative flux difference with respect to the point source pattern, when using different filters. The patterns correspond to an occultation such as (201) Penelope occulting TYC\,278-748-1 from \cite{natureAstronomy}, shown in figure \ref{fig:Penelope_lightcurves}, with an asteroid at a distance of 2.49 AU with a speed of about 2.2 km/s and a star size of 0.094 mas. VERITAS optical bandpass values correspond to the T1 telescope.
}

\end{figure}

\begin{figure}
\centering\includegraphics[width=0.9\linewidth]{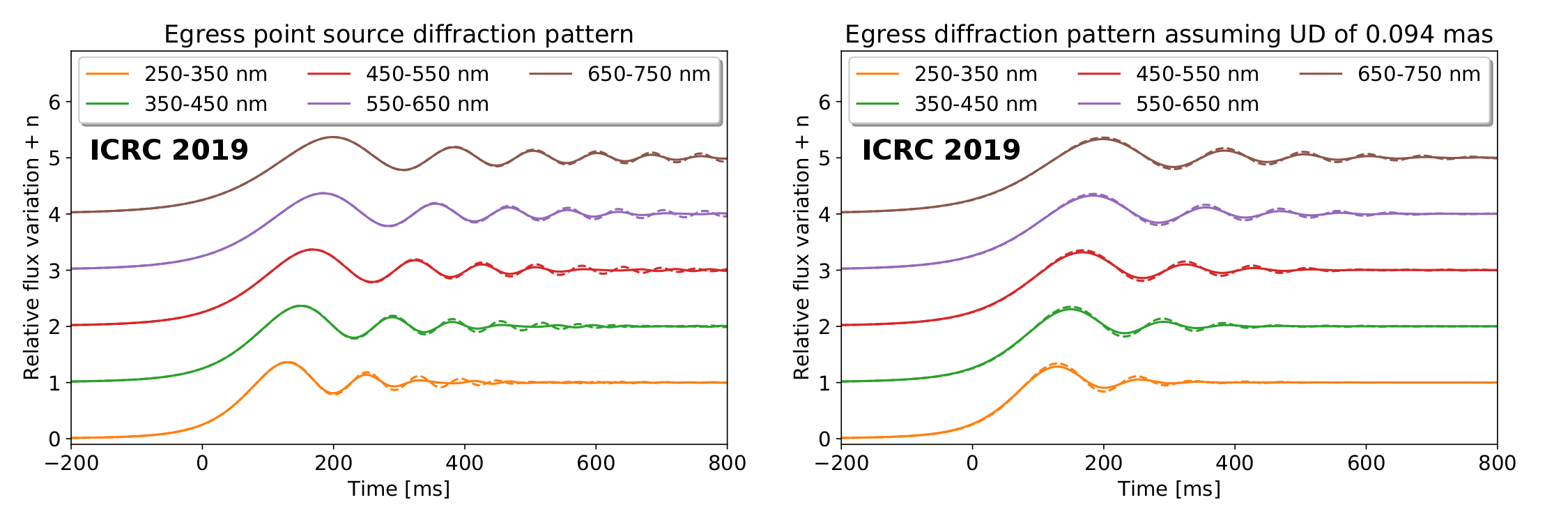}
\caption{\label{fig:multiwavelength_bandpass}
Diffraction pattern over different optical bandpass produced by a point-like source (left figure) and an extended uniform disk source (right figure). The optical bandpass used is a flat filter 100-nm wide (50 nm in case of the dashed lines). Same parameters used as in Figure \ref{fig:narrowing_bandpass}.
}

\end{figure}

Section \ref{sec:opticalVeritas} described the setup used for recent VERITAS optical measurements. Modifications to the hardware could be implemented to improve the instrument sensitivity and reduce the relative errors of measured stellar angular sizes. Here we list some examples of possible improvements: 
\begin{itemize}

\item \textbf{Using filters to narrow the optical bandpass}: As shown in Figure \ref{fig:narrowing_bandpass}, the contrast of the diffraction pattern produced by a point-like source is increased with a  narrower optical bandpass on the measured photons (about 10\% fringe amplitude in case of a 50 nm filter). When testing how such a narrower visible bandpass affects the observations reported within \cite{natureAstronomy}, that is by comparing the diffraction patterns produced by a point-source and a uniform disc of 94 micro-arcseconds, differences between both models increase by 5-7\% in amplitude.

\item \textbf{Simultaneous multiwavelength observations}: As discussed in section \ref{sec:opticalVeritas}, the sensitivity of IACTs are limited by the high night-sky background signal. Splitting the light beam to several visible wavebands would allow simultaneous multiwavelength measurements and would not significantly deteriorate sensitivity. This would also increase the number of independent measurements taken during every occultation. In addition, different wavelengths would produce significantly different diffraction patterns (as shown in figure \ref{fig:multiwavelength_bandpass}), improving the accuracy of the star size measurement. Note the size of the star is determined by the wavelength-dependent surface of last scattering of the photosphere, so both effects will be convoluted in the final measurement.

\item \textbf{Optimize the optical aperture}: as done in \cite{HESS_optical}, the night-sky background contribution (the main limitation of IACTs sensitivity in optical) can be reduced by stopping down the aperture for light entering the PMT Winston cones (usually hexagonal for IACTs). Sensitivity gains are the largest for the smallest ratio of the optical point spread function with respect to the Winston cone aperture.

\end{itemize}

\section{Conclusions}
\label{sec:conclusions}

In this work we present the asteroid occultation diffraction fitting technique as a viable method to expand the number of directly measured stellar diameters. As described, the different techniques that have been previously used to perform these measurements have various  limitations associated with the target stars detectable by each method, and limitations on achievable angular resolution. Asteroid occultations will allow to directly measure the size of fainter stars (easily down to the 13$^{th}$ magnitude) with sub-milliarcsecond resolution. One key aspect of the technique is the extremely short observation time that is required for these measurements. Similar to Lunar occultations, only a few minutes of observation are required for a quality measurement. In the case of asteroids, occultation events can be predicted in advance, allowing observatories to target only high-probability events.

As shown in Fig. \ref{fig:predicted_occultations} (left), the number of expected occultations detectable from a fixed site increases exponentially with the limiting magnitude. By using high-speed visible-band photometers with a sensitivity down to the 13$^{th}$ magnitude, such as IACTs, one could perform one measurement per week during the night time (not accounting for the impact of sky conditions such as full/partial Moon or weather). This rate is sufficient to obtain a viable sample of stellar radii for stellar population studies, useful for stellar evolution modelling \cite{Mozurkewich2003}.

The predicted number of asteroid occultations is especially interesting in the case of K stellar radii, as there is growing evidence of a discrepancy between stellar models and observations at the 5\% level \cite{Spada2013}.  Direct stellar radii measurement by interferometry is challenging, with about only 10 stars accessible to the current generation of interferometers over reasonable observation times. For stars brighter than 14$^{th}$ mag$_V$, 12\% of the expected asteroid occultations will be on K-type stars. This would mean that a single telescope with a sensitivity at the 13$^{th}$-magnitude level could double the amount of K stars with directly measured stellar diameters in a few years, while requiring a very small fraction of the telescope observation time (few minutes per week).

The usability of these measurements also depends on their precision. Even if relative errors better than 10\% need still to be demonstrated, several improvements have been proposed in section \ref{sec:improvements} that would help attaining such precision. Reaching $<$ 3\% relative error on stellar diameter measurements would make these observations impactful on the accurate determination of transiting exoplanet radii \cite{vonBraun2014}.

\section*{Acknowledgments}

This research is supported by grants from the U.S. Department of Energy Office of Science, the U.S. National Science Foundation and the Smithsonian Institution, and by NSERC in Canada. This research used resources provided by the Open Science Grid, which is supported by the National Science Foundation and the U.S. Department of Energy's Office of Science, and resources of the National Energy Research Scientific Computing Center (NERSC), a U.S. Department of Energy Office of Science User Facility operated under Contract No. DE-AC02-05CH11231. We acknowledge the excellent work of the technical support staff at the Fred Lawrence Whipple Observatory and at the collaborating institutions in the construction and operation of the instrument.

\end{document}